\documentclass[3p,times,pdflatex]{elsarticle}

\usepackage{ecrc}

\volume{00}

\firstpage{1}

\journalname{Nuclear Physics A}

\runauth{}

\jid{procs}

\jnltitlelogo{Nuclear Physics A}

\usepackage{amssymb}

\usepackage[figuresright]{rotating}

\usepackage{graphicx}
\usepackage{subfigure}
\usepackage{hyperref}

\begin{document}

\begin{frontmatter}



\dochead{}

\title{Quark splitting in non-trivial $\theta$-vacuum}


\author[CCNU,LBL]{Hongxi Xing}
\author[LBL]{Xin-Nian Wang}
\author[LBL,BNL]{Feng Yuan}
\address[CCNU]{Institute of Particle Physics and Key Laboratory of Quark {\&} Lepton Physics,
Huazhong Normal University, Wuhan 430079, China}
\address[LBL]{Nuclear Science Division, Lawrence Berkeley National Laboratory, 1 Cyclotron Road,
Berkeley, California 94720, USA}
\address[BNL]{RIKEN BNL Research Center, Building 510A, BNL, Upton, New York 11973, USA}

\begin{abstract}
Quark splitting in non-trivial $\theta$-vacuum with a given helicity
is investigated in pQCD with a modified quark propagator. We
found that the quark splitting functions were modified by the presence of
a topologically non-trivial QCD background field, though there is  no
explicit helicity flip associated with the radiative processes. The interaction with the
topological non-trivial field leads to the degeneracy of the quark
splitting functions for left- and right-handed quarks. Such
degeneracy can lead to imbalance of left- and right-handed quarks
in quark jet showers. We also discuss phenomenological consequences of such imbalance
if there exists non-trivial topological gluon field configuration in heavy-ion collisions.
\end{abstract}




\end{frontmatter}


\section{Introduction}
\label{} Quantum Chromodynamics (QCD) contains topologically
non-trivial configurations of gauge fields that can be represented
by degenerate vacuum states \cite{vacuum} and the physical
vacuum, so-called $\theta$-vacuum, could be a superposition of these
degenerate states.  The presence of the
$\theta$-vacuum state can be characterized by a $\theta$-term
in the QCD Lagrangian which would lead to
$\mathcal{CP}$ violation in the strong interaction. Search for the violation
of global $\mathcal{CP}$-invariance in strong interaction has only lead to an upper bound
on the value of $\theta$ from the neutron dipole moment, $\theta<3\times10^{-10}$
\cite{theta value}, indicating the absence of global $\mathcal{P}$ and
$\mathcal{CP}$ violation in QCD. Recently, the STAR Collaboration at
RHIC observed charge asymmetry in the azimuthal angle of dihadron correlation with respect
to the reaction plane in non-central heavy-ion collisions \cite{star}.
Such charge asymmetry is speculated to originate from the chiral magnetic
effect \cite{theta-dependence} due to the presence of local
$\mathcal{P}$ and $\mathcal{CP}$ violation in QCD at high temperature.
However, the asymmetry is found to exist both in and out of the
reaction plane which is so far not understood \cite{koch}.

\par In this talk, we present a study of parity-odd effect in processes of
parton shower in jet fragmentation by
considering the interaction of a quark with the non-trivial gauge
field in the $\theta$-vacuum. Assuming quarks propagate in the presence of a topological
non-trivial gluon field, we find a modified quark splitting probability that is
different for left- and right-handed quarks. QCD evolution equations with such modified
splitting functions lead to a sizable imbalance of shower quark distributions
for left- and right-handed quarks.


\section{Quark splitting in normal vacuum ($\theta=0$)}
\begin{figure}
\centering \subfigure[]{
\label{fig:subfig:a} 
\includegraphics[width=2in]{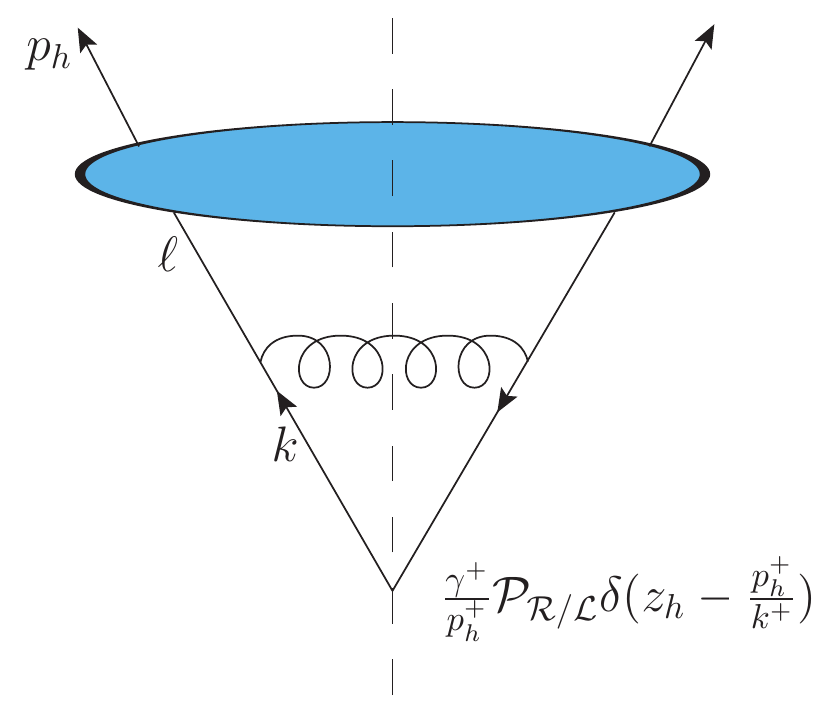}}
\hspace{1in} \subfigure[]{
\label{fig:subfig:b} 
\includegraphics[width=2in]{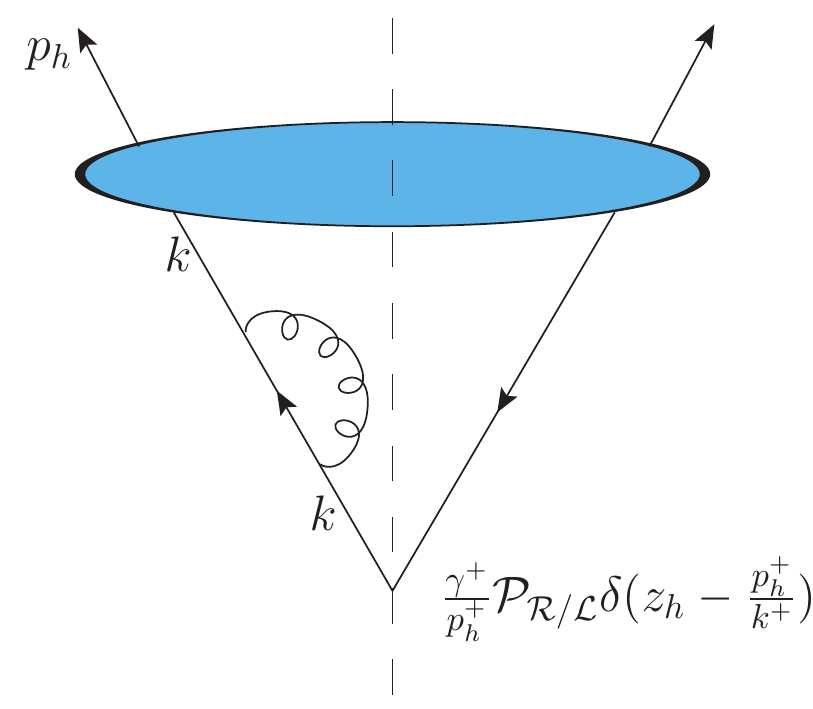}}
\caption{Diagrams for real (a) and virtual (b) corrections
contribute to quark fragmentation function.}
\label{fig:diagram} 
\end{figure}

In general, the left-handed quark fragmentation function is the same
as the right-handed  because of the parity invariance of the QCD
Lagrangian when $\theta=0$. In terms of parton matrix elements, it can
be defined as \cite{ffdefinition}:
\begin{eqnarray}
\label{ffdefine} D_{q_{L}\rightarrow h}(z_h)=D_{q_{R}\rightarrow
h}(z_h)=\frac{z_h}{2}\int{\frac{dy^-}{2\pi}e^{-ip_h^+y^-/z_h}
\sum_STr\left[\frac{\gamma^+}{2}\langle0|\psi(0)|p_h,S\rangle\langle
S,p_h|\psi(y^-)|0\rangle\right]},
\end{eqnarray}
where $D_{q_{L}\rightarrow h}(z_h)$ and $D_{q_{R}\rightarrow
h}(z_h)$ are left- and right-handed quark fragmentation functions,
respectively, with $z_h=p_h^+/k^+$ the momentum fraction of the quark carried
by the hadron. Here we use light-cone
notation $k^\pm=(k^0\pm k^z)/\sqrt{2}$. In normal vacuum, gluon bremsstrahlung or quark
splitting,  illustrated by the Feynman diagrams in Fig.~\ref{fig:diagram}, leads to the
Dokshitzer-Gribov-Lipatov-Altarelli-Parisi(DGLAP) \cite{DGLAP} QCD
evolution equations for quark fragmentation functions with given helicity,
\begin{eqnarray}
\label{DGLAP} \frac{\partial}{\partial{ln\mu^2}}\left(
\begin{array}{cc}
D_{q_{R}\rightarrow h}(z_h,\mu^2)\\
D_{q_{L}\rightarrow h}(z_h,\mu^2)
\end{array}
\right)= \frac{\alpha_s}{2\pi}\int_{z_h}^1{\frac{dz}{z}}\left(
\begin{array}{cc}
P_{q_Rq_R}(z) & P_{q_Lq_R}(z)\\
P_{q_Rq_L}(z) & P_{q_Lq_L}(z)
\end{array}
\right) \left(
\begin{array}{cc}
D_{q_{R}\rightarrow h}(z_h/z,\mu^2)\\
D_{q_{L}\rightarrow h}(z_h/z,\mu^2)
\end{array}
\right).
\end{eqnarray}
where we only consider the quark branching via gluon bremsstrahlung to analyze the
helicity distribution and  $P(z)$ are splitting functions:
\begin{eqnarray}
\nonumber
P_{q_Lq_L}(z)&=&P_{q_Rq_R}(z)=C_F\left[\frac{1+z^2}{(1-z)_+}+\frac{3}{2}\delta(z-1)\right],\\
P_{q_Rq_L}(z)&=&P_{q_Lq_R}(z)=0.
\end{eqnarray}
Suppose there are no differences between the
number of left- and right-handed quarks, or zero chirality,
before the start of branching processes,  the QCD evolution in normal
vacuum according to Eq.~(\ref{DGLAP}) does not induce any non-zero chirality.
This means that hadron distributions from left and right-handed quarks
should be identical. Therefore, alignment of left and right-handed quarks in the
presence of large magnetic field in heavy-ion collisions, does not lead to any
asymmetry in the final hadron spectra.  However, in the presence of a topological
non-trivial gluon field, the above evolution equation will be different for left and
right-handed quarks that might lead to hadron spectra asymmetry under a strong magnetic
field. This is essentially a dynamic mechanism for the proposed chiral magnetic
effect \cite{theta-dependence} .

\section{Quark splitting in non-trivial $\theta$-vacuum ($\bar{\theta}\neq0$)}
In QCD, the axial anomaly in quantum theory might lead to a
non-trivial $\theta$-vacuum which violates the parity conservation
in strong interaction. In this case, the vacuum wave function in QCD is
a linear combination of wave functions with different winding numbers.
This vacuum state can be reproduced by adding to the QCD
Lagrangian a new term,
\begin{equation}
\mathcal{L}_{\theta}=\theta/(32\pi^2)g^2F_{\mu\nu}^a\tilde{F}_a^{\mu\nu}.
\end{equation}
The observable effect of the above parity-violating interaction can
be mimicked by an effective space-time dependent dynamical "spurion"
field, i.e., $\theta=\theta(\textbf{x},t)$ \cite{theta-dependence}.
By performing an axial U(1) rotation and omit a full derivative
term, the extra $\theta$-term in QCD Lagrangian can be transformed
into the fermionic contribution $\mathcal{L}_{\theta}\rightarrow
1/(2N_f)\bar{\theta}_{\mu}\bar{\psi}\gamma^{\mu}\gamma^5\psi$
\cite{thetabar}, here $\bar{\theta}_{\mu}=\partial_{\mu}\theta$
represents the fluctuation of the $\theta$-vacuum, whose zero
component is the chiral chemical potential $\mu_5$. The existence of
this new term will yield a modified quark propagator
\cite{propagator}, in massless case,
\begin{eqnarray}
i\tilde{S}(k,\bar{\theta})=i\left[\mathcal{P_R}S(k+\bar{\theta})+
\mathcal{P_L}S(k-\bar{\theta})\right],
\end{eqnarray}
where  $\mathcal{P_{R/L}}$ are the right (left) projection operator
$\mathcal{P_{R/L}}=(1\pm\gamma^5)/2$, $iS(k)=i/\slash\!\!\!k$ is the
conventional quark propagator in normal vacuum.

With this modified quark propagator in non-trivial $\theta$-vacuum,
the parity symmetry will be broken in the process of quark propagation and
branching during quark fragmentation.  This parity-odd effect can be
described by the modified DGLAP evolution equation in $\theta$-vacuum,
\begin{eqnarray}
\label{mDGLAP}
\frac{\partial}{\partial{ln\mu^2}}\left(
\begin{array}{cc}
d\tilde{D}_{q_{R}\rightarrow h}(z_h,\mu^2)\\
d\tilde{D}_{q_{L}\rightarrow h}(z_h,\mu^2)
\end{array}
\right)= \frac{\alpha_s}{2\pi}\int_{z_h}^1{\frac{dz}{z}}\left(
\begin{array}{cc}
\tilde{P}_{q_Rq_R}(z) & \tilde{P}_{q_Lq_R}(z)\\
\tilde{P}_{q_Rq_L}(z) & \tilde{P}_{q_Lq_L}(z)
\end{array}
\right) \left(
\begin{array}{cc}
\tilde{D}_{q_{R}\rightarrow h}(z_h/z,\mu^2)\\
\tilde{D}_{q_{L}\rightarrow h}(z_h/z,\mu^2)
\end{array}
\right),
\end{eqnarray}
with the modified splitting functions,
\begin{eqnarray}
\label{msplitting} \nonumber
\tilde{P}_{q_Rq_R}(z,t)&=&C_F\left[\frac{1+z^2}{(1-z)_+}-t(1+t)\frac{1+z}{(z+t)^2}
+\left(\frac{3}{2}+a(t)\right)\delta(z-1)\right];\\
\nonumber
\tilde{P}_{q_Lq_L}(z,t)&=&C_F\left[\frac{1+z^2}{(1-z)_+}+t(1-t)\frac{1+z}{(z-t)^2}
+\left(\frac{3}{2}+a(-t)\right)\delta(z-1)\right];\\
\tilde{P}_{q_Rq_L}(z,t)&=&\tilde{P}_{q_Lq_R}(z,t)=0,
\end{eqnarray}
where $t=\bar{\theta}^+/k^+$, and
\begin{eqnarray}
\label{aterm}
a(t)=t\left[(1+t)\log{\frac{1+t}{z_v+t}}+(1-t)\frac{1-z_v}{z_v+t}\right],
\end{eqnarray}
$z_v$ is the minimum value of the momentum fraction bounded by the
virtuality of the initial quark. There are no helicity flip
because of the pseudovector quark-gluon coupling. However, the
parity is not conserved because of the degeneracy in the right-handed
splitting function $\tilde{P}_{q_Rq_R}(z,t)$ and left-handed splitting function
$\tilde{P}_{q_Lq_L}(z,t)$.

\par Compared to the DGLAP equation in normal vacuum, Eq.~(\ref{DGLAP}),
the renormalization equation for the modified quark fragmentation
function in non-trivial $\theta$-vacuum Eq. (\ref{mDGLAP}) is the
same as the evolution equation in normal vacuum except for the
modification of the splitting functions. As indicated in Eq.
(\ref{msplitting}), the modified splitting functions have an
extra term that is different for left- and right-handed quarks. It is these different
extra terms that lead to the parity-odd effect in quark branching processes for
non-vanishing  $\bar{\theta}$. One can see from Eq. (\ref{mDGLAP}),  the
DGLAP evolution equations in normal vacuum are recovered when $\bar{\theta}=0$.

\section{Shower parton distribution}
\begin{figure}
\centering
\includegraphics[width=5in]{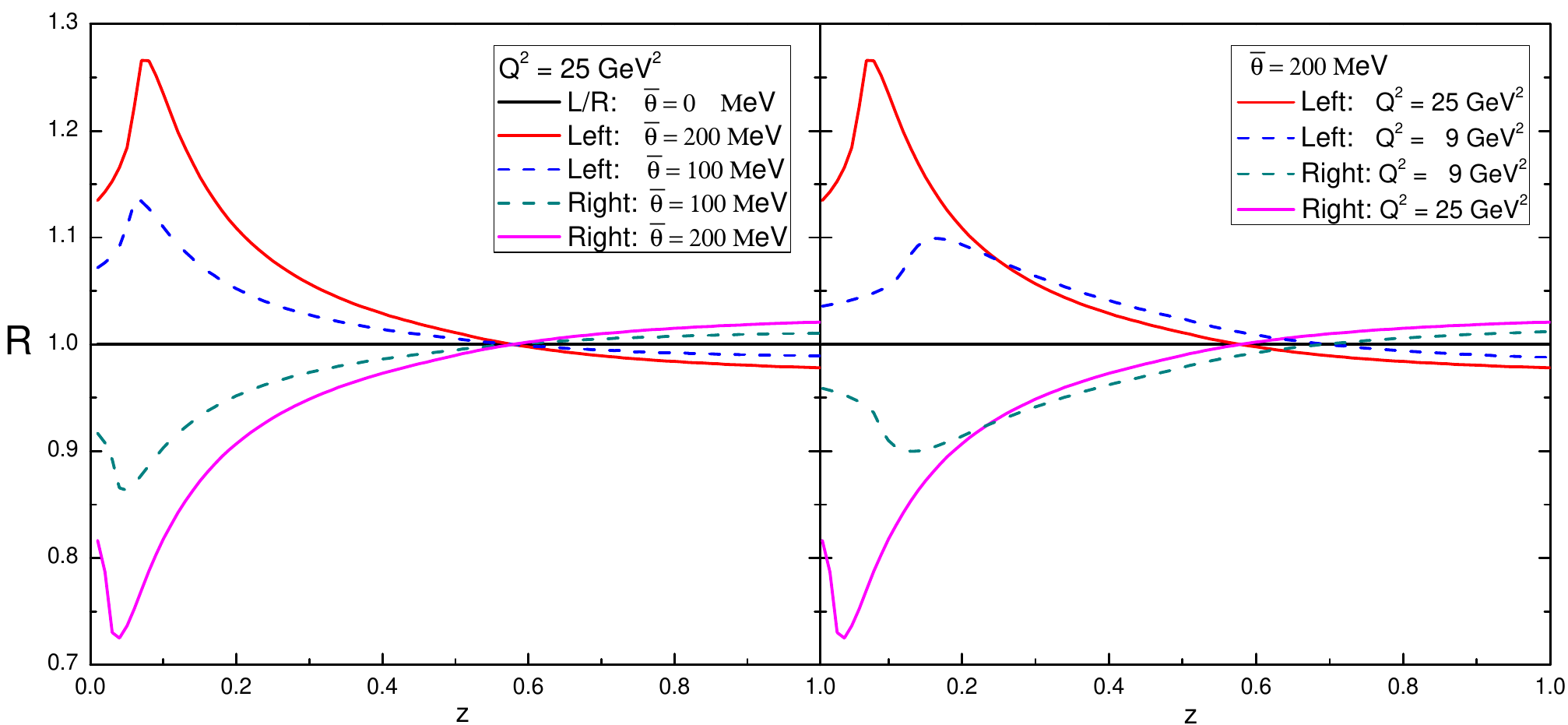}
\caption{Ratio defined in Eq. (\ref{Eq:ratio}) for different values
of $\bar{\theta}$ and momentum scale $Q^2$ for left- and
right-handed quarks.} \label{fig:ratio}
\end{figure}

To illustrate the imbalance of the left- and right-handed quark
distributions, we solve the modified DGLAP evolution equations
with an initial shower parton distribution,
\begin{eqnarray}
D_a^b(z,Q_0^2)=\delta_a^b\delta(z-1),
\end{eqnarray}
for both right and left-handed quarks. The index $a, b$ denote the quark flavors. To quantify the
effect of the $\theta$-vacuum, we define the ratio:
\begin{eqnarray}
\label{Eq:ratio} R \equiv
\frac{D_a^a(z,Q^2)|_{\bar{\theta}}}{D_a^a(z,Q^2)|_{\bar{\theta}=0}}.
\end{eqnarray}

\par Shown in Fig. \ref{fig:ratio}, are the ratios for
different values of $\bar{\theta}$ (left panel) and momentum scale
$Q^2$ (right panel) as a function of the momentum fraction $z$. One
can see that the interaction with the topological non-trivial gauge
field configuration lead to the degeneracy of left- and right-handed
quark distributions. This degeneracy or imbalance between the left-
and right-handed quark distributions is proportional to the value of
$\bar{\theta}$.  Therefore, a non-zero chirality proportional to the
$\theta$-vacuum fluctuation was generated by the modified DGLAP
evolution equations.

In the right panel of Fig. \ref{fig:ratio}, we also observe that the imbalance of
right and left-handed quark distributions or chirality
from the modified evolution equations is mostly in the region of small and moderate momentum
fraction $z$ and the degeneracy increases with  momentum scale $Q^{2}$. In principle, one can
obtain the final state charged hadron distribution by
convolution of this shower quark distribution function with the
normal fragmentation function, assuming quark hadronization below scale $Q_{0}$ happens
in normal vacuum.


\par We have illustrated that the parity-odd terms in the modified DGLAP evolution equations
lead to a net helicity for quarks in the parity-odd domain, which is proportional to the
fluctuation of the $\theta$-vacuum. Under a strong magnetic
field, the left and right-handed quarks would orient their spin parallel (anti-parallel) to the
magnetic field due to electromagnetic interaction through their magnetic moments.
The above net chirality would then lead to hadron or charge asymmetry with respect to the
direction of the magnetic field. Therefore, a combination of topological non-trivial gluon
field and strong magnetic field would lead to hadron asymmetry in the final state
of quark fragmentation. In high-energy heavy-ion collisions, strong magnetic field
is induced by two high-energy nuclei with finite impact-parameter. Such magnetic field
exists only for a very short period of time during the early stage of heavy-ion collisions, during
which hard processes and jet parton branching happen. Therefore, existence of
non-trivial gluon field would then lead to hadron asymmetry in the final hadrons from
jet fragmentation.

\section{Summary}
In this paper, we have studied quark splitting in a
non-trivial QCD vacuum with a given helicity. We found that the
quark splitting functions were modified differently for left- and
right-handed quarks for quark branching under such topologically non-trivial QCD
background. This difference will induce non-zero chirality. In our
calculation, this chirality is not induced by the flipping of the
quark helicity, but rather by the branching asymmetry for left- and
right-handed quarks in the presence of $\theta$-vacuum. The
modifications of the splitting functions are parity-odd and depend
on the size of the $\theta$ fluctuation. We have estimated this
parity violation effect in the quark
distribution in the parton shower of a quark jet. We found
a sizable imbalance of quark distribution for left- and
right-handed quarks. Such imbalance could provide an mechanism for
final hadron asymmetry induced by the $\theta$-vacuum fluctuation.

The authors thank Zhong-Bo Kang for helpful discussion. This work is
supported  by DOE under Contract No. DE-AC02-05CH11231 and NSFC
under Project Nos. 10525523, Nos. 10825523 and Nos. 10635020.



\bibliographystyle{elsarticle-num}



\end{document}